# Pressure pulsations in reciprocating pump piping systems Part 1: modelling

**Jian-Jun SHU** [a], **Clifford R. BURROWS** [b] and **Kevin A EDGE** [b]
[a] School of Mechanical & Aerospace Engineering, Nanyang Technological University, 50 Nanyang Avenue, Singapore 639798
[b] Fluid Power Centre, School of Mechanical Engineering, University of Bath

**Abstract**: A distributed parameter model of pipeline transmission line behaviour is presented, based on a Galerkin method incorporating frequency-dependent friction. This is readily interfaced to an existing model of the pumping dynamics of a plunger pump to allow time-domain simulations of pipeline pressure pulsations in both suction and delivery lines. A new model for the pump inlet manifold is also proposed.

## 1 INTRODUCTION1

Reciprocating plunger pumps are robust, contamination tolerant and capable of efficiently pumping many types of fluids at high delivery pressures. As a consequence, they are widely used in a diverse range of industrial applications, including mining (for powered roof supports), chemical plant, reverse osmosis systems and food processing systems. The most common pump construction consists of a small number of cylinders, usually mounted in-line, each with a reciprocating piston driven by a rotating crank and connecting rod mechanism. During the suction stroke, flow is drawn from the inlet manifold into a cylinder through a self-acting non-return valve; various valve designs are employed although spring-load poppet or disc valves are most frequently adopted. Fluid delivery also takes place through a self-acting non-return valve.

It is well known that the pipeline pressure pulsations produced by these pumps are a source of noise and vibration and may have a significant influence on the reliability of a given installation. Consequently, it is highly desirable to be able to predict pressure pulsations at the design stage of an installation so that appropriate steps may be taken to minimize their levels and their influence.

Considerable research effort has been devoted to the study of pressure pulsation behaviour in delivery lines of fluid power systems employing typically gear, vane or axial piston pumps [e.g. see references (**1**) and (**2**)] and to a lesser extent to the suction lines of these systems (**3**, **4**). However, fluid power pumps typically employ a large number of pumping elements (nine cylinders are commonly used in axial piston machines, for example). As a consequence they create relatively low amplitude flow pulsations and, in most instances, low-amplitude pressure pulsations with a relatively high frequency content are generated. This allows a linearized analysis to be adopted and predictions can be conveniently conducted in the frequency domain.

In contrast, plunger pumps generally have a small number of cylinders (three or five are common) and are usually operated at lower speeds. This leads to very large flow pulsations, relative to the mean flow. It is possible that the consequent high-amplitude pressure pulsations, particularly in resonant delivery line systems, may invalidate the use of linear theory. Hence predictions of behaviour need to be performed either in the time domain or by means of an iterative scheme [e.g. see reference (**5**)]. A frequency domain approach to the prediction of suction line pulsation behaviour is likely to be invalid if cavitation is occurring as the effects are highly non-linear.

Some useful progress has already been made by a number of workers on the mathematical modelling of the pumping dynamics of reciprocating plunger pumps. Johnston (**6**), for example, has developed a detailed model which accounts for both valve dynamics and cavitation in the pump cylinders. However, inlet line pressure is taken to be constant and the delivery line is represented by a lumped parameter model. Vetter and Schweinfurter (**7**) address the problems of delivery pipeline wave propagation effects, but adopt a fairly rudimentary pump model. Most of their predictions of pulsation behaviour are presented in terms of peak-to-peak pulsation levels, rather than frequency spectra or time-domain waveforms. Thus the accuracy of the model is difficult to establish. Singh and Madavan (**5**) have presented a more detailed model of pumping dynamics which is linked to a frequency-domain model of the delivery pipeline. An iterative process is used to account for the interactions between the pipeline and the pump. Predicted pressure pulsation behaviour is compared with experimental data in terms of amplitude spectra. Phase spectra are not included in the paper, so, again, it is difficult to establish the accuracy of the model in predicting behaviour. Vetter and Schweinfurter do not attempt to predict suction line pulsations and although Singh and Madavan claim that their model will predict suction line behaviour, no results are presented.

This paper aims to address the inadequacy of existing models by describing the development of a finite element model of pipeline dynamics (under non-cavitating and cavitating conditions) which is integrated with an existing generic model of pumping dynamics.

## 2 PUMP MODEL

This study concentrates on a single-acting, in-line plunger pump, one cylinder of which is shown schematically in Fig. 1. A detailed model for such an arrangement has been developed by Johnston (**6**). The model accounts for flow continuity into and out of each cylinder, according to whether the inlet or delivery valve is open. Inlet pressure is taken to be constant. In Johnston's approach the flows from cylinders on their delivery stroke are summed together and the resultant is used as the input to a lumped parameter model of the delivery line. Full account is taken of the forces acting on the inlet and delivery valves to provide comprehensive modelling of valve dynamics. Full details of the model are given in reference (**6**). It should be noted that the approach is readily adapted to suit other pump configurations.

The modifications necessary for interfacing Johnston's pump model to distributed parameter models of suction and delivery lines will now be presented.

### 2.1 Manifold modelling

The flows from those cylinders communicating with the delivery manifold are assumed to be created at one discrete location rather than being spatially distributed over the length of the manifold. This simplifies the interfacing of the pump model with the delivery line model. It would not be difficult to develop a spatially distributed model of flows into the delivery manifold but the agreement between predictions and experimental data suggests that this is an unnecessary refinement. The summed flows are introduced at the internal end of the manifold and define the boundary conditions for the delivery line model (Section 3). The manifold itself is treated as part of the delivery line.

Experience has shown that a more detailed model may be required for the inlet manifold. Once again, the flows relating to individual cylinders communicating with the manifold are all assumed to occur at the same location. However, it has been argued that as a result of air release, air pockets can form in the inlet manifold. These, combined with pressure losses, play an important role in the suction dynamics.

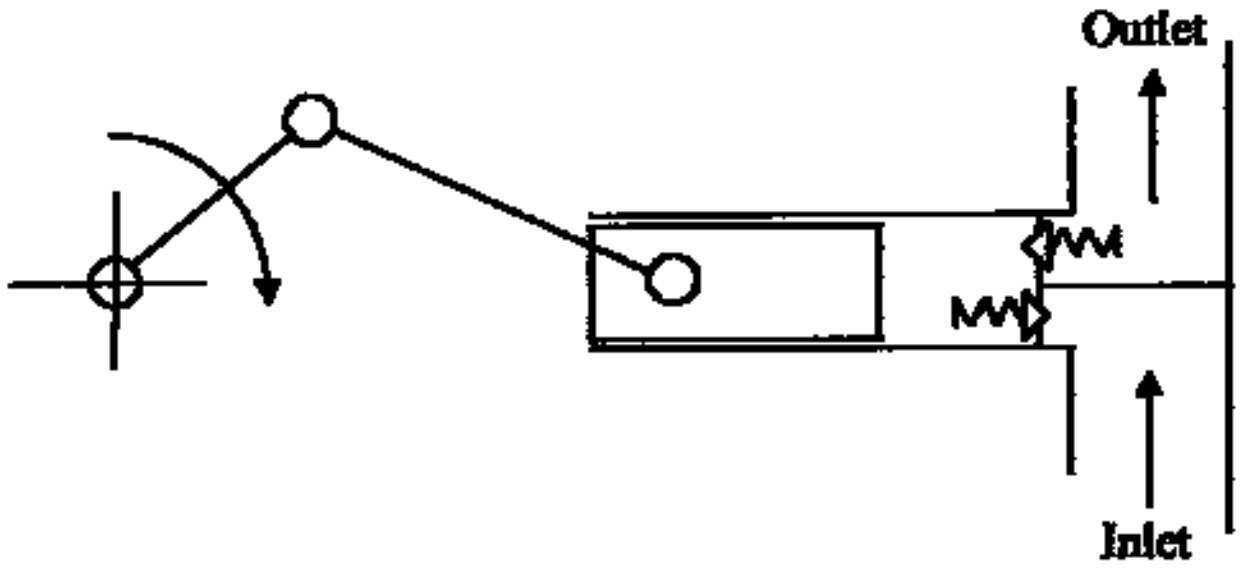

**Fig. 1** Schematic of one cylinder of a reciprocating pump

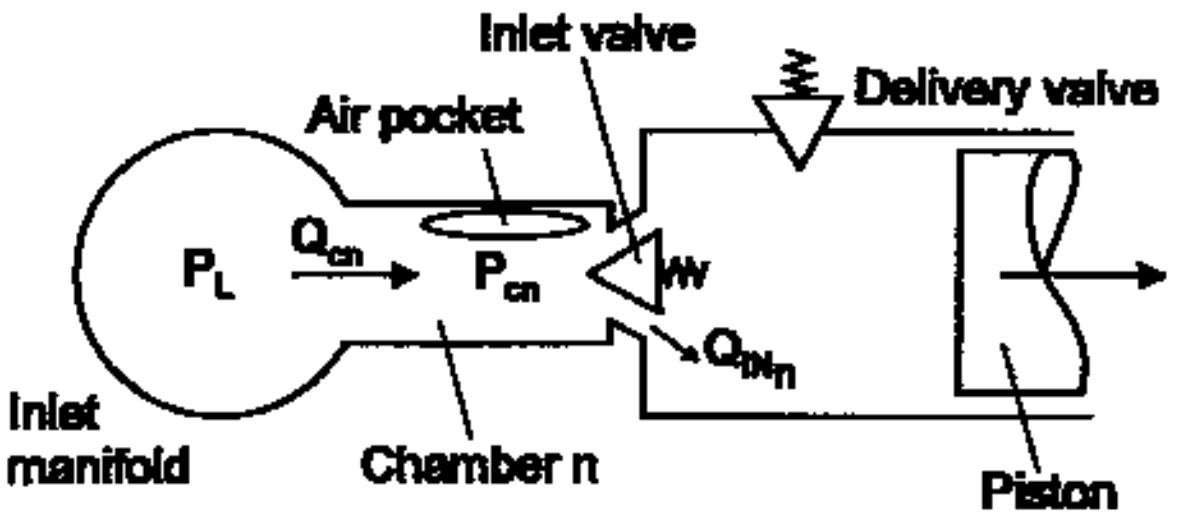

**Fig. 2** Schematic of inlet chamber geometry used in the mathematical model

To model such behaviour a small chamber is assumed to be present upstream of each inlet valve, as illustrated schematically in Fig. 2. Each chamber can contain a pocket of air and communicates with the inlet manifold via a square-law restrictor. This is close to many real pump designs where the inlet valve is located at the end of a (usually short) passageway at right angles to the manifold. Pressure losses are introduced by the right-angled bend and, depending on manifold geometry, air pockets could be trapped near the inlet valve. Through the selection of appropriate restrictor loss coefficients, some account can be taken of the different pressure losses likely to be experienced at different locations along the manifold.

The relevant equations for this model are

$$Q_{cn} = K_{cn}\sqrt{(P_L - P_{cn})} \quad (1)$$

and

$$\frac{dP_{cn}}{dt} = \frac{B_{cn}}{V_{cn}}(Q_{cn} - Q_{INn}) \quad (2)$$

Numerical integration of equation (2) gives the chamber pressures, which on substitution in equation (1) provide the flows drawn from the inlet manifold. The total flow drawn from the inlet line is

$$Q_c = \sum_{n=1}^{M} Q_{cn} \quad (3)$$

That part of the inlet manifold not incorporated in chamber models is included as part of the inlet line.

To account for the presence of air pockets in the chambers, an effective bulk modulus of elasticity is required. From reference (**8**),

$$\frac{B_{cn}}{B} = \frac{V_{1n}/V_{an} + P_0/P_{cn}}{V_{1n}/V_{an} + P_0 B/P_{cn}^2} \quad (4)$$

In this instance, $p_0 B/p_{cn}^2 \gg V_{1n}/V_{an}$ and hence

$$B_{cn} = P_{cn}\left(1 + \frac{V_{1n}P_{cn}}{V_{an}P_0}\right) \quad (5)$$

The principal problem with this detailed model is the difficulty in selecting appropriate values for the volumes $V_{ln}$, $V_{an}$ and the pressure loss coefficient $K_{cn}$. This may be eased somewhat by selecting the same parameters for each restrictor and chamber combination, albeit at the expense of losing the ability to model cylinder-to-cylinder variations. However, if this approach is acceptable, significant gains in computational efficiency can be achieved by assuming the existence of just one chamber with which all cylinders communicate. This chamber is linked to the manifold through one square-law restrictor. In the case of a three-cylinder pump the error introduced has been found to be acceptably small since the 'overlap' between two cylinders communicating with the chamber at the same time is small, relative to the period of each suction stroke.

## 3 TRANSMISSION LINE MODELLING

The dynamics of distributed parameter piping systems are described by hyperbolic partial differential equations. A commonly used numerical scheme to solve these equations is the method of characteristics (**9**-**11**) which has been widely and successfully employed to model fluid transient behaviour such as waterhammer under non-cavitating and cavitating conditions. However, because the spatial discretization of the line is intrinsically linked to the time step and speed of sound in the fluid, difficulties can be encountered in obtaining compatibility with the small time steps required to solve the differential equations describing components connected to the line (**12**). For example, for a time step of $10^{-2}$ ms and a speed of sound of 1000 m/s, a line 10 m long would need to be divided into 1000 elements. Moreover, when variable time steps are required, the calculation of intermediate values by interpolation becomes a further computational burden.

To avoid these problems, an alternative approach is adopted in this study in which the Galerkin finite element method (**13**, **14**) is applied in the spatial variables only. This gives rise to an initial value problem for a system of ordinary differential equations, allowing the time step to be decoupled from the spatial interval.

The flow within a transmission line is to be calculated under the assumptions of one-dimensional, unsteady compressible flow. Independent variables of space and time are denoted by x and t. The dependent variables are the pressure P and the flowrate Q. Hence two partial differential equations have to be solved:

Equation of continuity:

$$\frac{1}{c_0^2}\frac{\partial P}{\partial t} + \frac{\rho}{\pi r_0^2}\frac{\partial Q}{\partial x} = 0 \quad (6)$$

Equation of motion:

$$\frac{\rho}{\pi r_0^2}\frac{\partial Q}{\partial t} + \frac{\partial P}{\partial x} + F(Q) + \rho g \sin\theta_0 = 0 \quad (7)$$

In the case of laminar flow the friction term F(Q) can be expressed as a quasi steady term $F_0$ plus an unsteady term (`frequency-dependent friction'), for which an approximation has been developed by Zielke (**15**) and Kagawa *et al*. (**16**):

$$F(Q) \approx F_0 + \frac{1}{2}\sum_{i=1}^{k} Y_i \quad (8)$$

where

$$F_0 = \frac{8\mu Q}{\pi r_0^4} \quad (9)$$

and

$$\frac{\partial Y_i}{\partial t} = -\frac{n_i \mu}{\rho r_0^2} Y_i + m_i \frac{\partial F_0}{\partial t}$$

$$Y_i(0) = 0 \quad (10)$$

The constants $n_i$ and $m_i$ are given by Kagawa *et al*. (**16**) and are reproduced in Table 1. The number of terms k should be selected according to the frequency range of interest.

**Table 1** Values of $n_i$ and $m_i$ for use in equation (10)

| $i$ | $n_i$ | $m_i$ |
|---|---|---|
| 1 | $2.63744 \times 10^1$ | 1.0 |
| 2 | $7.28033 \times 10^1$ | 1.16725 |
| 3 | $1.87424 \times 10^2$ | 2.20064 |
| 4 | $5.36626 \times 10^2$ | 3.92861 |
| 5 | $1.57060 \times 10^3$ | 6.78788 |
| 6 | $4.61813 \times 10^3$ | $1.16761 \times 10^1$ |
| 7 | $1.36011 \times 10^4$ | $2.00612 \times 10^1$ |
| 8 | $4.00825 \times 10^4$ | $3.44541 \times 10^1$ |
| 9 | $1.18153 \times 10^5$ | $5.91642 \times 10^1$ |
| 10 | $3.48316 \times 10^5$ | $1.01590 \times 10^2$ |

For turbulent flow, the quasi-steady term is replaced by

$$\frac{\rho f |Q| Q}{4\pi^2 r_0^5}$$

The unsteady friction term developed for laminar flow [equation (10)] has been found by Vardy *et al.* (**17**) to work well at Reynolds numbers up to about $10^4$ and has been adopted here. For cases involving much higher Reynolds numbers the model proposed by Vardy *et al.* (**17**) could be adopted.

### 3.1 Galerkin finite element method

Finite element formulations based on the Galerkin method (**18**) for time domain analysis have been presented by Rachford and Ramsey (**13**) and Paygude *et al.* (**14**) using a conventional uniformly spaced grid system with two degrees of freedom (pressure and flowrate). In the work that follows the method is presented for one degree of freedom (either pressure or flowrate), with an extension to include the effects of frequency-dependent friction. The calculation of pressure and flow at alternating nodes is sometimes referred to as interlacing.

Equations (6) to (10) can be rearranged in terms of operator equations:

$$L_1(U, P, Y_i) \equiv \frac{1}{c_0^2}\frac{\partial P}{\partial t} + \frac{\partial U}{\partial x} = 0 \qquad (11)$$

$$L_2(U, P, Y_i) \equiv \frac{\partial U}{\partial t} + \frac{\partial P}{\partial x} + R'U + \frac{1}{2}\sum_{i=1}^{k} Y_i + H_0 = 0 \qquad (12)$$

$$L_3(U, P, Y_i) \equiv \frac{\partial Y_i}{\partial t} + \frac{n_i R}{8} Y_i - m_i R \frac{\partial U}{\partial t} = 0 \qquad (13)$$

where

$$U = \frac{\rho Q}{\pi r_0^2}, \qquad R = \frac{8\mu}{\rho r_0^2} \qquad \text{and} \qquad H_0 = \rho g \sin\theta_0$$

For laminar flow, $R' = R$. For turbulent flow, the quasi-steady term may be approximated by

$$R' = \frac{f|U_{-1}|}{4\rho r_0} \qquad (14)$$

where $U_{-1}$ is the value for U from the previous time step. The transmission line is divided into 2N+1 equal elements, each Δx in length. A minimum of five elements is required.

The Galerkin method involves finding approximations to U, P and $Y_i$ of the form

$$U(x, t) = u_{2j}(t)w_{2j}^{+}(x) + u_{2j+2}(t)w_{2j+2}^{-}(x) \qquad (15)$$

$$P(x, t) = p_{2j+1}(t)w_{2j+1}^{+}(x) + p_{2j+3}(t)w_{2j+3}^{-}(x) \qquad (16)$$

$$Y_i(x, t) = y_{i,2j}(t)w_{2j}^{+}(x) + y_{i,2j+2}(t)w_{2j+2}^{-}(x) \qquad (17)$$

for $j = 0, 1, \ldots, N-1$. The unknown coefficients u, p and y are nodal values of U, P and $Y_i$ respectively.

The weighting (or basis) functions $w^+$, $w^-$ are piecewise polynomials. Here, linear interpolation functions are adopted:

$$w_m^{+}(x) = \begin{cases} \dfrac{x_{m+2} - x}{x_{m+2} - x_m} & \text{for } x_m \leqslant x \leqslant x_{m+2} \\ 0 & \text{otherwise} \end{cases} \qquad (18)$$

$$w_m^{-}(x) = \begin{cases} 1 - w_m^{+}(x) & \text{for } x_m \leqslant x \leqslant x_{m+2} \\ 0 & \text{otherwise} \end{cases} \qquad (19)$$

Nodal values are determined by inner products $\langle .,. \rangle$ defined as follows:

$$(L_1, w_{2j+1}^{+}) \equiv \int_{x_{2j+1}}^{x_{2j+3}} w_{2j+1}^{+}(x) L_1 \, dx = 0 \qquad (20)$$

$$(L_1, w_{2j+1}^{-}) \equiv \int_{x_{2j+1}}^{x_{2j+3}} w_{2j+1}^{-}(x) L_1 \, dx = 0 \qquad (21)$$

and, for $e = 2$ and $e = 3$,

$$(L_e, w_{2j}^{+}) \equiv \int_{x_{2j}}^{x_{2j+2}} w_{2j}^{+}(x) L_e \, dx = 0 \qquad (22)$$

$$(L_e, w_{2j}^{-}) \equiv \int_{x_{2j}}^{x_{2j+2}} w_{2j}^{-}(x) L_e \, dx = 0 \qquad (23)$$

Evaluation of these integrals results in a set of ordinary differential equations which allow the calculation of the pressures at the odd-numbered nodes and flows at the even-numbered nodes. In this study it was decided to specify the flow at one end of the line and the pressure at the other end (although it is equally possible to formulate solutions for the cases where either pressure is defined at both ends or flow is defined at both ends). In order to establish each boundary condition it is necessary to solve the equation relating to the end condition simultaneously within the equation describing the behaviour of the attached component. This necessitates the use of a different weighting function, spanning a single element rather than two, for the elements at each end of the line. The resultant grid is illustrated in Fig. 3.

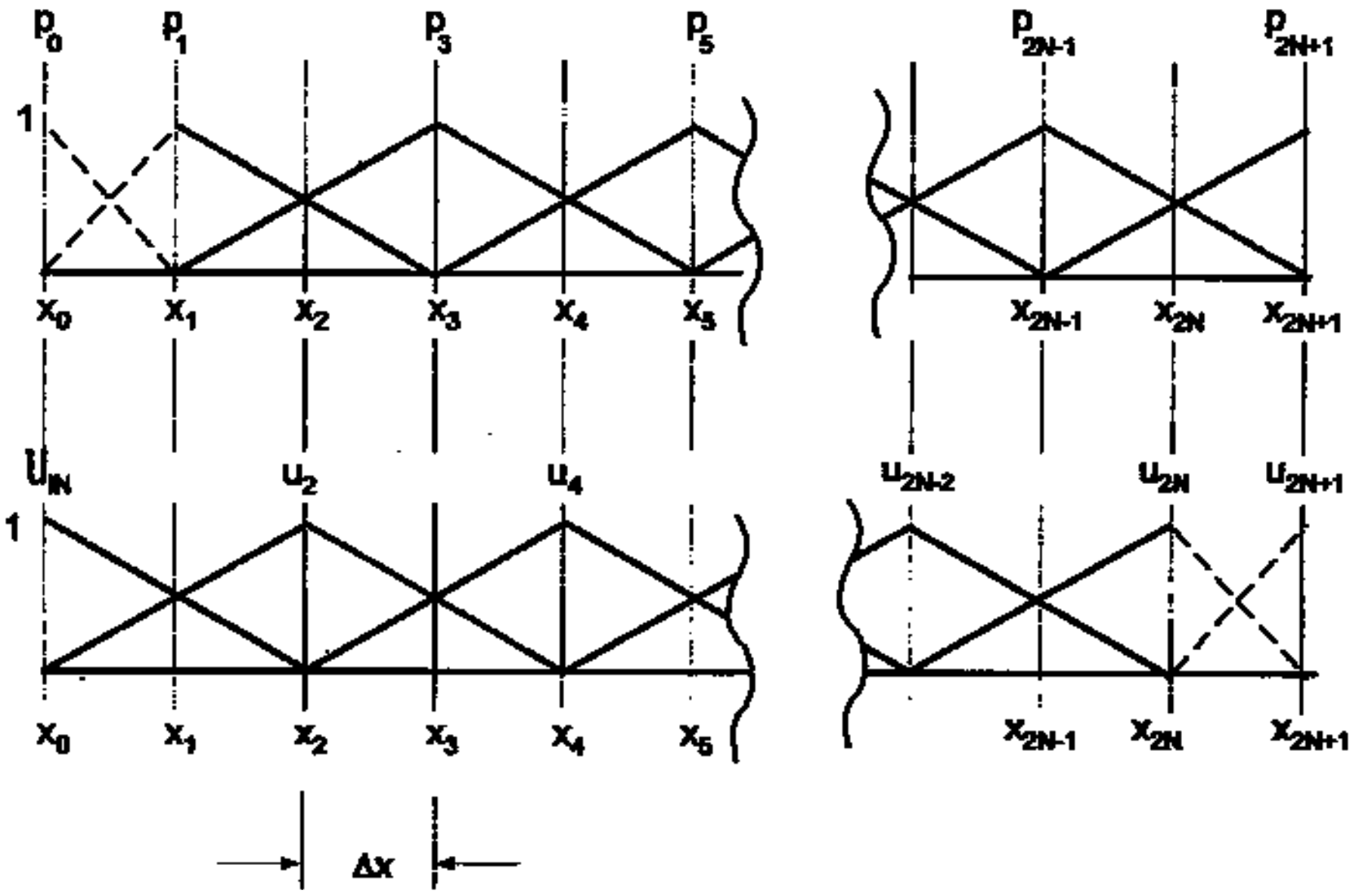

**Fig. 3** Pipeline discretization grid and weighting functions

For the interlaced nodes, the resultant ordinary differential equations are of the following form:

$$\frac{\mathrm{d}}{\mathrm{d}t}\begin{pmatrix} \boldsymbol{u} \\ \boldsymbol{p} \\ \boldsymbol{y} \end{pmatrix} = \begin{pmatrix} -R'\mathbf{I} & \frac{1}{16\Delta x}\mathbf{A} & -\frac{1}{2}(\Xi \otimes \mathbf{I})^{\mathrm{T}} \\ \frac{c_0^2}{16\Delta x}\mathbf{B} & \mathbf{0} & \mathbf{0} \\ -R^2\mathbf{M}\otimes\mathbf{I} & \frac{R}{16\Delta x}\mathbf{M}\otimes\mathbf{A} & -\frac{R}{8}[\mathbf{F} + 4\mathbf{M}\otimes(\Xi\otimes\mathbf{I})^{\mathrm{T}}] \end{pmatrix} \times \begin{pmatrix} \boldsymbol{u} \\ \boldsymbol{p} \\ \boldsymbol{y} \end{pmatrix} + \frac{c_0^2}{16\Delta x}\mathrm{U_{IN}}\begin{pmatrix} \mathbf{0} \\ \mathbf{C} \\ \mathbf{0} \end{pmatrix} + \frac{1}{16\Delta x}P_{\mathrm{OUT}} \times \begin{pmatrix} \mathbf{D} \\ \mathbf{0} \\ R\mathbf{M}\otimes\mathbf{D} \end{pmatrix} - H_0\begin{pmatrix} \mathbf{E} \\ \mathbf{0} \\ R\mathbf{M}\otimes\mathbf{E} \end{pmatrix} \quad (24)$$

The symbol $\otimes$ defines the Kronecker product of a vector $\boldsymbol{Z} = \{z_1, z_2, \ldots, z_k\}^{\mathrm{T}}$ and a matrix $\mathbf{G}$, i.e.

$$\boldsymbol{Z}\otimes\mathbf{G} \equiv \begin{pmatrix} z_1\mathbf{G} \\ z_2\mathbf{G} \\ \vdots \\ z_k\mathbf{G} \end{pmatrix} \quad (25)$$

where

$$\boldsymbol{u} = \{u_2, u_4, \ldots, u_{2N}\}^{\mathrm{T}}$$

$$\boldsymbol{p} = \{p_1, p_3, \ldots, p_{2N-1}\}^{\mathrm{T}}$$

$$\boldsymbol{y} = \{y_{1,2}, y_{1,4}, \ldots, y_{1,2N}, y_{2,2}, \ldots, y_{k,2N}\}^{\mathrm{T}}$$

$$\mathbf{N} = \{n_1, n_2, \ldots, n_k\}^{\mathrm{T}}, \qquad \mathbf{M} = \{m_1, m_2, \ldots, m_k\}^{\mathrm{T}}$$

$$\Xi = \{1, 1, \ldots, 1\}^{\mathrm{T}}$$

$$\mathbf{A} = \begin{pmatrix} 10 & -12 & 2 & & & & & \\ -1 & 11 & -11 & 1 & & & & \\ & -1 & 11 & -11 & 1 & & & \\ & & \cdot & \cdot & \cdot & \cdot & & \\ & & & \cdot & \cdot & \cdot & \cdot & \\ & & & & \cdot & \cdot & \cdot & \cdot \\ & & & & -1 & 11 & -11 & 1 \\ & & & & & -1 & 11 & -11 \\ & & & & & & -2 & 12 \end{pmatrix}$$

$$\mathbf{B} = \begin{pmatrix} -12 & 2 & & & & & & \\ 11 & -11 & 1 & & & & & \\ -1 & 11 & -11 & 1 & & & & \\ & \cdot & \cdot & \cdot & \cdot & & & \\ & & \cdot & \cdot & \cdot & \cdot & & \\ & & & \cdot & \cdot & \cdot & \cdot & \\ & & & & -1 & 11 & -11 & 1 \\ & & & & & -1 & 11 & -11 & 1 \\ & & & & & & -2 & 12 & -10 \end{pmatrix}$$

$$\mathbf{C} = \{10, -1, 0, \ldots, 0, 0, 0\}^{\mathrm{T}}$$

$$\mathbf{D} = \{0, 0, 0, \ldots, 0, 1, -10\}^{\mathrm{T}}$$

$$\mathbf{E} = \{1, 1, 1, \ldots, 1, 1, 1\}^{\mathrm{T}}$$

$$\mathbf{F} = \begin{pmatrix} n_1\mathbf{I} & & & & \\ & n_2\mathbf{I} & & & \\ & & \cdot & & \\ & & & \cdot & \\ & & & & \cdot \\ & & & & & n_k\mathbf{I} \end{pmatrix}$$

The nodal values at the boundary conditions are obtained by applying the same methodology but using different weighting functions, as dictated by the grid shown in Fig. 3. This results in two further ordinary differential equations:

$$\frac{\mathrm{d}p_0}{\mathrm{d}t} = -\frac{1}{2}\frac{\mathrm{d}p_1}{\mathrm{d}t} + \frac{3c_0^2}{4\Delta x}(u_0 - u_2) \tag{26}$$

(where $u_0 \equiv U_{\mathrm{IN}}$) and

$$\frac{\mathrm{d}u_{2N+1}}{\mathrm{d}t} = -\frac{1}{2}\frac{\mathrm{d}u_{2N}}{\mathrm{d}t} + \frac{3}{4\Delta x}(p_{2N-1} - p_{2N+1}) - \frac{R'}{2}(u_{2N} - 2u_{2N+1}) - \frac{3}{2}H_0 \tag{27}$$

(where $p_{2N+1} \equiv P_{OUT}$). For the case of the pump inlet line, the boundary condition at the reservoir is the constant pressure source, $p_{2N+1}$. At the pump, the total instantaneous flow drawn into the cylinders defines $U_{IN}$. However, individual cylinder flows are dictated by the instantaneous inlet valve differential pressure. Hence, equation (26) must be used to provide $p_0$, thereby allowing the flows to be calculated.

At the pump outlet, the total delivery flow is established using the same procedure, which now provides $U_{IN}$ for the delivery line. In the case of a valve terminating the delivery line, equation (26) is solved simultaneously with the equation describing the valve behaviour. For a simple restrictor valve returning the flow to atmospheric pressure, for example, the termination pressure would be described by the square law relationship:

$$p_{2N+1} = c_V Q_{LV}^2 \tag{28}$$

where $Q_{LV}$ is established from $u_{2N+1}$ at the previous time step. This approach may be extended to include valve dynamics if required.

To accommodate the possibility of cavitation in the suction line, three different pipeline cavitation models were considered, based on the work of Shinada and Kojima (**19**). A vaporous cavitation model was adopted, following its successful implementation in a previous investigation employing the method of characteristics (**11**). In this scheme the pressure at any node is constrained not to fall below the vapour pressure. Further details of the approach are given in reference (**11**).

## 4 CONCLUSIONS

It has been argued that current models of plunger pumps are inadequate in respect of the complex interactions which take place between the pump and attached pipelines. These arise because of the distributed parameter nature of the pipelines and because of cavitation. A finite difference method for modelling pipelines, based on a Galerkin method incorporating frequency-dependent friction, has been proposed. This approach circumvents the computationally intensive demands associated with the use of the method of characteristics.

A new model for the pump inlet manifold has been developed to account for the presence of air pockets. The pipeline models are readily interfaced to an existing model of pumping dynamics, to allow time-domain simulations of pressure pulsations.